\documentclass[preprint,12pt]{elsarticle}

\usepackage{amssymb}
\usepackage{bm}
\usepackage{amsmath}
\usepackage{float}
\usepackage{hhline} 
\usepackage{booktabs}
\usepackage{color,ulem}
\usepackage{hyperref}

\hypersetup{
	colorlinks=true,    
	linkcolor=blue,     
	citecolor=blue,    
	urlcolor=red       
}

\journal{Physica A: Statistical Mechanics and its Applications}

\begin{document}

\begin{frontmatter}



\title{Structural dynamics of a model of amorphous silicon}


\author[inst1]{Zihua Liu}
\author[inst1]{Debabrata Panja}
\author[inst1]{Gerard T. Barkema}

\affiliation[inst1]{organization={Department of Information and Computing Sciences, Utrecht University},
            addressline={ Princetonplein 5 }, 
            city={Utrecht},
            postcode={3584 CC}, 
            country={The Netherlands}}
            
\begin{abstract}
We perform extensive simulations and systematic statistical analyses on the structural dynamics of a model of amorphous silicon. The simulations follow the dynamics introduced by Wooten, Winer and Weaire: the energy is obtained with the Keating potential, and the
dynamics consists of bond transpositions proposed at random
locations and accepted with the Metropolis acceptance ratio. The
structural quantities we track are the variations in time of the
lateral lengths ($L_x$, $L_y$, $L_z$) of the cuboid simulation cell.
We transform these quantities into the volume $V$ and two aspect
ratios $B_1$ and $B_2$. Our analysis reveals that at short times,
the mean squared displacement (MSD) for all of them exhibits normal
diffusion. At longer times, they cross over to anomalous
diffusion, with a temperature-dependent anomalous
exponent $\alpha<1$. We analyze our findings in the light of two
standard models in statistical physics that feature anomalous
dynamics, {\it viz.\/}, continuous time random walker (CTRW) and
fractional Brownian motion (fBm). We obtain the distribution of
waiting times, and find that the data are consistent with a
stretched-exponential decay. We also show that the three
quantities, $V$, $B_1$ and $B_2$ exhibit negative velocity
autocorrelation functions. These observations together suggest that
the dynamics of the material belong to the fBm class.

\end{abstract}



\begin{keyword}
Structural dynamics \sep Amorphous silicon  \sep Monte Carlo method \sep Stochastic process
\PACS 0000 \sep 1111
\MSC 0000 \sep 1111
\end{keyword}

\end{frontmatter}

\section{Introduction\label{sec1}}

Amorphous silicon (a-Si) has attracted a lot of attention in the
computational materials science community over the last decades,
partly because it is a material with many applications, and partly
because it has become a prototypical example of a covalently bonded
disordered material that lends itself well for simulations, as many
empirical and semi-empirical potentials are readily available.  There
exist a myriad of experimental and simulation research works on the
properties of a-Si, studying thermal transport \cite{braun2016size,zhou2020thermal}, structure and defects recognition \cite{tielens2020characterization}, electronic properties and applications for solar cells \cite{ru202025,cao2020light,huang2021comparative}, and 
vibrational properties \cite{nguyen2020abrasive}. 
Compared to structural and electronic properties, the dynamics of a-Si
is less studied. In this paper we fill the knowledge gap on how structural properties of the material evolve in time. Dynamics of covalently-bonded materials are well-known to be extremely slow. On the one hand, it makes them highly stable, but on the other it renders them nearly impossible to be accessed by experimental time scales. Computer simulations can provide a way forward, but there too, simulation studies of a-Si dynamics have often focused on how to generate well-relaxed samples as efficiently as possible, but do not study the fundamental properties of its dynamics.

In this paper, we employ a relatively simple model for the dynamics of
a-Si, as first introduced by Wooten, Winer and Weaire (WWW)
\cite{wooten1985computer}. The atomic structure in this model is
described as a so-called continuous random network (CRN), in which
each silicon atom has exactly four covalent bonds to other silicon
neighbors. To each CRN-configuration, an energy is attributed using
the Keating potential \cite{keating1966effect}.  The dynamics consists
of a sequence of bond transpositions proposed at random locations, and
accepted or rejected according to the Metropolis algorithm. This technique allows for the simulation of the dynamics of a-Si over time scales that are much longer than those accessible to molecular dynamics (MD), at the expense of being a much cruder description. There is a however a connection between the two time scales. Within our Monte Carlo (MC) dynamics, the probability that a specific bond 
	transposition is proposed in one MC unit of time is
	$2/(4*3*3*N)=1/(18N)$. Here, factors $1/N$, $1/4$, $1/3$ and $1/3$
	respectively arise from picking a random atom (out of $N$ total
	atoms), then one of its four neighbours, next twice one of the three
	remaining neighbors, and the factor of $2$ comes from the possibility
	to generate the same string of atoms from two different ends.
	The acceptance probability is then $\exp(-\beta \Delta E)$, where
	$\Delta E$ is the energy difference between the initial and final states. Thus,
	the rate of structural changes in the sample is
	$(1/18N)\exp(-\beta\Delta E)$.
	
	In molecular dynamics (MD), the rate would be $\nu\exp(-\beta B)$
	where $\nu$ is the attempt frequency, often found to be around
	$10^{-12}$ s$^{-1}$=1 ps$^{-1}$, and $B$ is the energy barrier
	between the initial and final states. The energy barrier $B$ and the
	energy change $\Delta E$ are correlated, but loosely. For instance, we know
	that $B$ has to be higher than $\Delta E$. For example, from earlier work
	[Ref. \cite{barkema1998identification}, Table 1] we know that on
	average for bond transpositions (WWW events),
	$\langle\Delta E\rangle=4.0$ eV and $\langle B\rangle=2.2$ eV. As a
	rough estimate of $\Delta E-B$, we can then take
	$\langle B\rangle - \langle\Delta E\rangle = 1.8$ eV. Our simulations
	were carried out at a temperature of 0.14 eV, with $N=2000$ atoms. We
	then obtain that one unit of MC time corresponds roughly to
	$(1/(18N)) \exp(1.8/0.14)$ ps $\approx$ 10 ps. 
	
In our simulations, we use a cuboid simulation cell with periodic
boundary conditions, and lateral dimensions $L_x\times L_y\times
L_z$. At all times, these lateral dimensions, together with all the atoms, constitute $N+3$ degrees of freedom. After each WWW move, the bond list is updated, and all degrees of freedom are minimized to the global minimum of energy, and consequently these lateral dimensions of the simulation cell will fluctuate in time. In our case, these fluctuations are easily accessible in simulations, and are closely related to the mechanical deformations under (small) external forces, and therefore a standard approach for understanding the mechanical shear and stress properties of the bulk material. All dynamics studied in the paper is performed at the temperature below the melting temperature of a-Si (1750K\cite{kluge1987amorphous}). 

We transform the three quantities $L_x(t)$, $L_y(t)$ and $L_z(t)$ into
three other quantities, which are the volume
$V(t) \equiv L_x(t) \cdot L_y(t) \cdot L_z(t)$ and the two aspect
ratios $B_1(t)=L_y(t)/L_z(t)$ and
$B_2(t)=L_y(t)\cdot L_z(t)/L_x^2(t)$.  The aspect ratios $B_1(t)$ and
$B_2(t)$ are two (almost) unconstrained degrees of freedom, which can
vary over a wide range of values without preference.  This however does not hold for the volume $V(t)$ as it is
constrained to fluctuate around an ideal value set by the number of
atoms and the ideal density.  (It is well-known that well-relaxed
samples of a-Si can be obtained within a wide range of densities:
fluctuations of several percent are easily observed. Thus, the
constraint is rather loose for this specific material.)

The focus of this paper lies on the dynamics of $V$, $B_1$ and
$B_2$. The motivation for this choice is that these are quantities
that lend themselves well for computer simulations studies, and are
directly connected to the mechanical behavior under shear, stress,
etc., which are of experimental relevance. Our findings indicate that over short time scales, all
three quantities exhibit diffusive behavior; this is to be expected,
as over a short time, the dynamics consist of local atomic
rearrangements which cause random changes in $B_1$, $B_2$ and $V$
which are uncorrelated. At longer times, a negative velocity
autocorrelation emerges: a positive change induces a bias at later
times to be followed by a negative change, vice versa. These
correlations are strong enough to change the dynamics from ordinary to
anomalous diffusion: the mean squared deviations MSD$_1$, MSD$_2$ and
MSD$_V$ for $B_1$, $B_2$ and $V$ can be fitted by a power-law
$\sim t^\alpha$ with $\alpha<1$. The exponent is found to be
temperature-dependent.  (Given that $V(t)$ is constrained to fluctuate
around an ideal value as noted above would mean that the mean-square
displacement of $V$ must reach a plateau at a long time. Our
simulation times are however not long enough, i.e., the mean-square
displacement of $V$ does not reach a high enough value to be
influenced by the constraint.)

We analyze our findings in the light of standard models for anomalous
diffusion, {\it viz.\/}, the continuous time random walk (CTRW) model
and fractional Brownian motion (fBm). For this purpose, we also
determine numerically the distribution of the waiting times.  This
distribution can be well fitted by a function with
stretched-exponential decay. Although this is different from homogeneous, memoryless materials, where the decay is
expected to be exponential, the decay is still too steep to be the
sole explanation of anomalous diffusion.  We also determine the
velocity autocorrelation functions, and find them to be negative at
all times.

The organization of this paper is as follows. In Sec. \ref{model}, we
introduce the model and simulation methods in detail. In
Sec. \ref{ini}, we define the structural quantities related to the 
mechanical properties, analyze the trajectories extracted from simulations,
investigate the short-time behavior of mean-square displacements, and
describe the relationships among the different diffusion
coefficients. In Sec. \ref{pro}, we analyze the probability
distribution, both on waiting time and jump length, characterize
sub-diffusive behaviors at long times, and reflect on the
classification of the model. In Sec. \ref{con} we summarize and
conclude the paper.

\section{The model \label{model}} 

\begin{figure}[h]
	\includegraphics[width=0.8\linewidth]{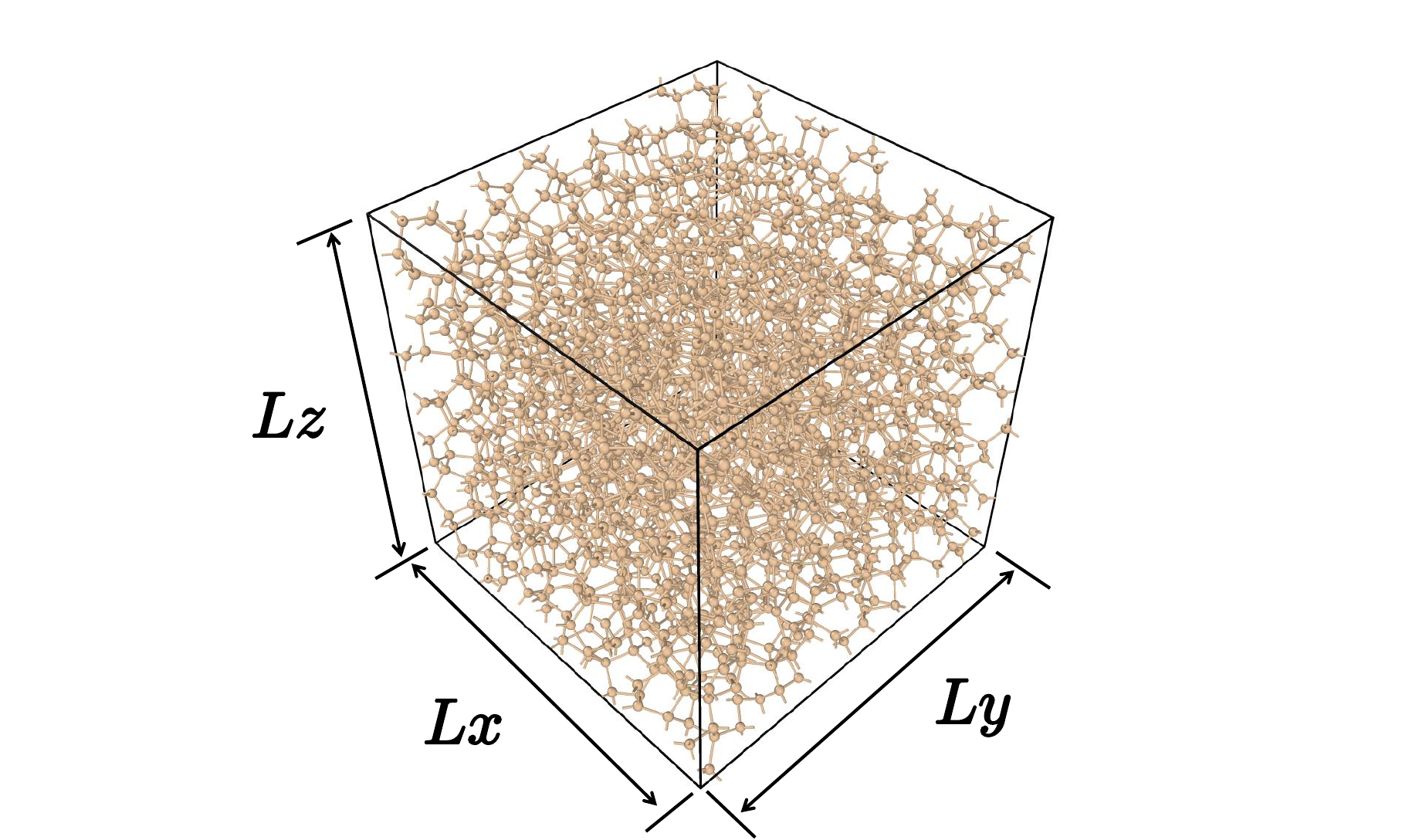}
	\centering
	\caption{(color online) An initial sample of amorphous silicon
		with 2000 atoms, each connected to four neighboring
		atoms. The bonds have comparable length (2.35\AA \space with a
		spread of 0.085\AA), and the bond angles are close to the
		tetrahedral angle ($109.46^\circ$ with a spread of
		$9.5^\circ$). The sample has been generated, starting from a
		Voronoi diagram and then evolved using the WWW-algorithm, as
		discussed in Sec. \ref{model}. It is a cuboid with periodic
		boundaries on the lateral dimensions $L_x$, $L_y$ and
		$L_z$. \label{fig1}}
\end{figure}

The Keating potential is one of the simplest and most efficient models
for describing a-Si. It uses an explicit list of bonds: whether two
atoms interact with each other or not, is thus determined by this bond
list, and not by the distance between the atoms. It is defined as
\begin{eqnarray}
	E &=& \frac{3}{{16}}\frac{\alpha }{{{d^2}}}\sum\limits_{ < ij > } 
	{({\textbf{r}_{ij}} \cdot {\textbf{r}_{ij}}}  - {d^2}{)^2} + \frac{3}{8}\frac{\beta }{{{d^2}}}\sum\limits_{ < ijk > } 
	{({\textbf{r}_{ij}} \cdot {\textbf{r}_{ik}} + \frac{1}{3}} {d^2}{)^2}.
	\label{eff-potential}
\end{eqnarray}
Here, $\textbf{r}_{ij}$ is the bond vector between atoms $i$ and
$j$. $d = 2.35$ \AA~is the ideal bond length for pure crystal silicon. $\alpha$ is two-body term constant, taken to be $2.965$eV\AA$/d^2$, $\beta$ is the three-body term, set as 0.285$\alpha$.

At the beginning of each simulation, an initial sample has to be
prepared.  In order to ensure that our initial sample is homogeneous
and isotropic, we start with randomly placed points in a cubic box,
and then determine the Voronoi diagram between these random
points. The set of edges in the Voronoi diagram forms a
fourfold-coordinated CRN, and each edge is then seen as a bond of the
initial explicit bond list. At the locations where four edges meet, a
silicon atom is placed. From this moment on, the initial random points
no longer have a role.

\begin{figure}[H]
	\begin{center}
		\includegraphics[width=0.8\linewidth]{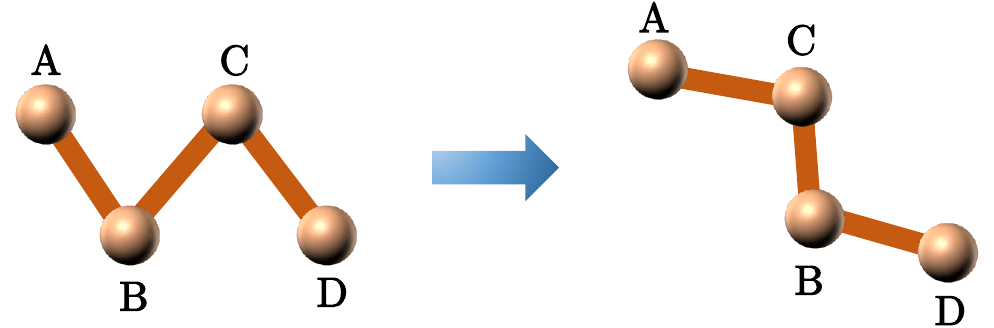}
	\end{center}
	\caption{(color online) {Diagram of a bond
			transposition which is part of the WWW algorithm. A string of
			connected atoms ABCD is chosen, then the bonds AB and CD
			are replaced by AC and BD. The new string is subsequently
			relaxed by the following local energy
			minimization.} \label{btrans}}
\end{figure}

Next, the atomic positions are allowed to relax, while preserving the
explicit bond list; this is done by a straightforward local energy
minimization (implemented as discussed below). The resulting initial
sample is homogeneous and isotropic, but poorly relaxed. An often used
characterization of the degree of relaxation is via the spread in bond
angles. While in experimental a-Si samples, the angular spread ranges
from 10 degrees for well-relaxed samples to 12 degrees for poorly
relaxed samples, these initial Voronoi-created samples have an angular
spread of 15 degrees or more. In order to relax the sample, we follow
the procedure initially proposed by Wooten, Winer and Weaire (WWW).
Randomly, somewhere in the sample, a string of four connected atoms
ABCD is chosen. Next, a bond transposition is made: the bonds
connecting these four atoms are reconnected, by replacing the bonds AB
and CD by bonds AC and BD, as shown in Fig. \ref{btrans}. This is
followed by a local energy minimization \cite{d2021efficient} under
the new explicit bond list, which changes the atomic coordinates
slightly.  This proposed bond transposition is then either accepted or
rejected, according to the Metropolis criterion
\cite{metropolis1953equation}. The acceptance probability is given by
\begin{equation}
	P_{\text{acc}}={\text{Min}} \left[1, \exp(-\beta \Delta E)\right],
\end{equation}
where $\Delta E$ is the change in energy due to the proposed bond
transposition, and $\beta=(k_{\text B} T)^{-1}$ with temperature $T$ and
Boltzmann constant $k_{\text B}$.  Typically, many thousands of such bond
transpositions are required, for obtaining a well-relaxed a-Si sample.

The time-consuming part of this WWW algorithm is the local energy
minimization. Our algorithm of choice for doing this, is the fast
inertial relaxation engine (FIRE) algorithm.  Typical parameters are
set as same as in the Ref. \cite{guenole2020assessment}: 
$N_{\textrm{min}}=5$,
$f_{\textrm{inc}}=1.1$, $f_{\textrm{dec}}=0.5$,
$\alpha_{\textrm{start}}=0.1$ and $f_{\alpha}=0.99$. Other custom
parameters here are set as $\Delta t_{\textrm{MD}}=0.06$,
$\Delta t_{\textrm{max}}\sim10\Delta t_{\textrm{MD}}$ and the velocity
Verlet method is chosen for integration in time.

To improve the computational efficiency, we use early rejection of
`hopeless' moves, as discussed in Ref. 
\cite{barkema2000high,d2021efficient,liu2022structural}. 
At each proposed bond
transposition, a threshold value for the energy is set. If the energy
after relaxation stays above the threshold, the proposed move is
rejected, otherwise it is surely accepted.  During energy minimization
after a proposed bond transposition, the total force is monitored, and
used to make a conservative estimate of the energy that would be
obtained after full minimization. If it is clear that this energy
stays above the threshold, the bond transposition is rejected well
before the time-costly full relaxation is achieved. In well-relaxed
samples, this early-rejection gives a speed-up of one or two orders of
magnitude.

\section{Dynamics of fluctuations in sample shapes\label{sec2}}

As shown in Fig. \ref{fig1}, we perform our simulations in the
periodic and cuboid simulation box $L_x$$\times$$L_y$$\times$$L_z$ in
the $x$-, $y$- and the $z$-directions respectively. These are however
not fixed quantities, but are allowed fluctuate when the bond
transpositions are made as the sample evolves in time.

Throughout this paper, we consider three geometric quantities, defined as
follows:
\begin{eqnarray}
	V(t) &=& L_x(t)L_y(t)L_z(t),\nonumber\\
	B_1(t) &=& L_y(t)/L_z(t),\nonumber\\
	B_2(t) &=& L_y(t)L_z(t)/L_x^2(t).
	\label{eVB1B2}
\end{eqnarray}

\begin{figure}[H]
	\begin{center}
		\includegraphics[width=\linewidth]{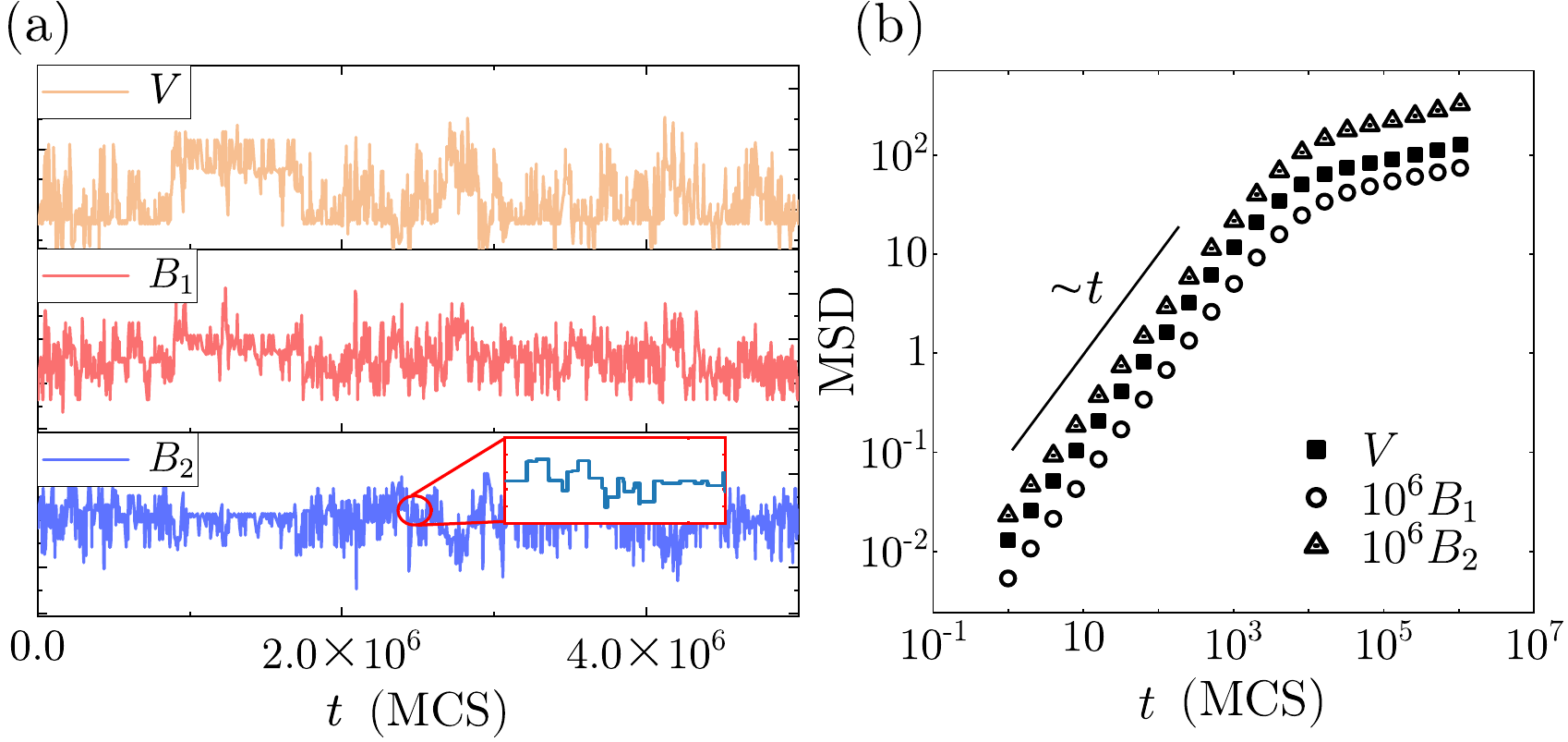}
	\end{center}
	\caption{(color online) (a) Typical fluctuations of $V$,
		$B_1$ and $B_2$ (from top to bottom) evolved
		up to 5$\times10^6$ Monte Carlo steps (MCS) for a sample with 
		$N=423$ 
		atoms under $T=1400$K.  While, given enough time, $B_1$ and $B_2$ 
		can 
		take a wide range of values, $V$ will only show fluctuations around 
		its
		ideal value which is set by the density of the a-Si sample. The
		inset in the bottom panel exhibits stalling events (see text for
		details). (b)  Measurements of the MSD for $V$, 
		$B_1$ and $B_2$ respectively . $B_1$ and $B_2$ are multiplied by 
		$10^6$ 
		to have the same scale as $V$. The results are obtained from 
		averaging 
		over 10 starting samples with 423 atoms ($T=1400$K),  
		 each evolved 50 
			times with different random seeds over $10^7$ proposed bond 
			transpositions to achieve a good statistical performance, which
			offset the noise generated by a single evolution. At short 
		times, all 
		three curves of the MSD initially increase linear in time, i.e. 
		show 
		ordinary diffusion. After $\sim 10^4$ attempted bond 
		transpositions, the
		increase slows down significantly. All curves show a crossover to
		subdiffusive behavior. For the MSD$_V$ we expect eventually
		saturation; this seems to be at times beyond those of our
		simulations.\label{fig2} }
\end{figure}

Physically, for a cuboid sample such as ours, the dynamics of shape
fluctuations of the sample can be efficiently characterized by
associating $V(t)$ to the dynamics of the ``bulk'' mode, and $B_1(t)$
and $B_2(t)$ with that of the ``shear'' modes. Fig. \ref{fig2}(a) shows
the trajectories for these quantities in Monte-Carlo time at short
times. The stalling events seen therein (inset, Fig.  {\ref{fig2}(a))
	result from the {\it rejected\/} bond transpositions at short time
	intervals. Figure \ref{fig2}(b) displays the MSD 
		for $V$, $B_1$ and $B_2$, respectively. At short times, 
		there are only few bond transpositions which occur at spatially 
		separated 
		locations, and thus these are uncorrelated events, which yield 
		normal diffusive 
		behavior. At longer times, this spatial separation breaks down, and 
		the bond 
		transpositions start to feel each other; for instance, a bond 
		transposition 
		which leads to local contraction has an enhanced probability to be 
		followed by 
		another bond transposition that leads to local expansion. The 
		cross-over point 
		represents the transition from normal diffusion to anomalous 
		diffusion of the 
		physical quantity under study.

	\subsection{Diffusive behavior at short times\label{ini}}
	
	At short times, we track the dynamics of shape fluctuations of the
	samples in terms of the mean-square displacements
	MSD$_V(t)=\langle[V(t)-V(0)]^2\rangle$,
	MSD$_1(t)=\langle[B_1(t)-B_1(0)]^2\rangle$ and
	MSD$_2(t)=\langle[B_2(t)-B_2(0)]^2\rangle$, with the angular
	brackets denoting ensemble averages for a sample of fixed number of
	atoms and (more or less) constant energy. As shown in
	Fig. \ref{fig2}(b), At short times, all three quantities exhibit
	ordinary diffusive behavior, i.e.  MSD($t$)$\sim t$. 
	$V$ is expected to saturate after super long times due to the
	constraint of structural density, while significant crossovers
	(nearly at $10^4$) to sub-diffusion can be observed in MSD$_1(t)$
	and MSD$_2(t)$ i.e. MSD($t$) $\sim t^\alpha$ ($\alpha < 1$). Here
	the results are obtained from averaging 10 starting samples
	($N=423$) each evolved 50 times.
	\begin{equation}
		\mbox{MSD}(t) = 2 Dt.
		\label{eq3}
	\end{equation}
	By fitting the MSD for $V$, $B_1$ and $B_2$ we extract the
	corresponding diffusion coefficients $D$. Since $V$, $B_1$ and $B_2$ all bear relations to $L_x$, $L_y$ and $L_z$, one would expect them to
	be related through these length parameters, which we establish
	below. In order to do so, having denoted the change in $V$, $B_1$ and
	$B_2$ over a small time interval $dt$ for samples with dimensions
	$L_x$ and $L_y$ by $dV$, $dB_1$ and $dB_2$ respectively, we express
	them in terms of small changes $dL_x$, $dL_y$ and $dL_z$ as
	\begin{eqnarray}
		\langle dV^2 \rangle\!\!\! &=& \!\!\!\left\langle\!\left[L_y L_z 
		dL_x + L_z
		L_x dL_y + L_x L_y dL_z\right]^2\right\rangle
		\nonumber\\
		\langle dB^2_1 \rangle\!\!\! 
		&=&\!\!\!\left\langle\!\frac{1}{L_z^4}\!\left[L_z
		dL_y - L_y dL_z\right]^2\!\right\rangle
		\quad\mbox{and}\nonumber\\
		\langle dB^2_2 \rangle\!\!\! 
		&=&\!\!\!\left\langle\!\frac{1}{L_x^6}\!\left[L_xL_ydL_z+L_zL_xdL_y
		- 2L_yL_zdL_x\right]^2\!\right\rangle\!\!.
		\label{eq4}
	\end{eqnarray}
	
	 Numerically, we find that cross-terms
		such as $\langle dL_xdL_y\rangle$ etc. are much smaller than terms
		like $\langle dL_x^2 \rangle$.  Similarly, it can be argued that
		thermal kicking on a sample acts like a force in the $x$-, $y$- and
		$z$- directions as stretching forces, and the sample cannot
		distinguish among the three directions. In particular, if the
		condition holds that the extension of the sample along the
		$x$-direction is inversely proportional to the corresponding spring
		constant $(L_yL_z)^{-1}$, then
		$\langle dL_x^2\rangle \sim (L_y L_z)^{-2}$ etc. These two
		assumptions lead to the following scaling relations between the
		diffusion coefficients:
		\begin{eqnarray}
			& D_V/D_{B_1}\propto\left\langle {L_x^2L_z^4} \right\rangle 
			\nonumber\\
			& D_{B_1}/D_{B_2}\propto\left\langle {L_x^4/L_z^4} 
			\right\rangle\nonumber\\
			& D_V/D_{B_2}\propto\left\langle {L_x^6} \!\right\rangle
			\label{rD}.
		\end{eqnarray}
		
		\begin{figure}[H]
			\begin{center}
				\includegraphics[width=\linewidth]{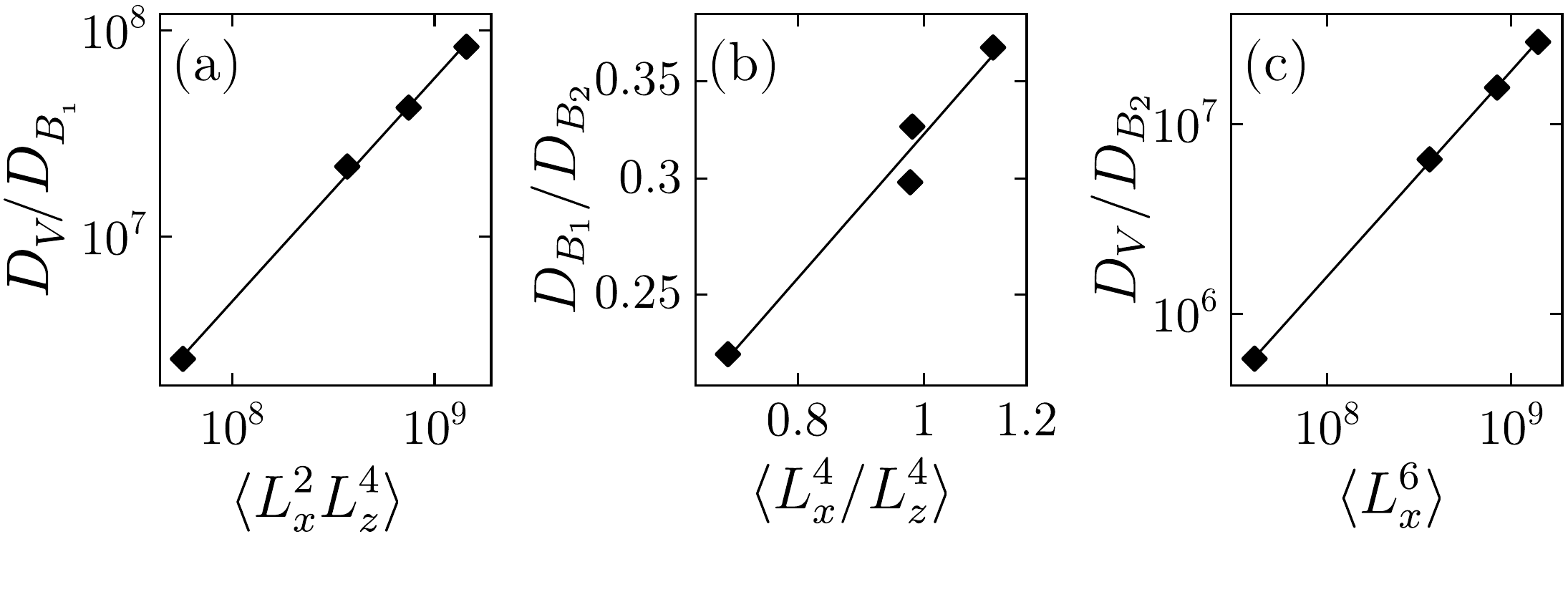}
			\end{center}
			\caption{(a)-(c) Double logarithmic plots showing the
				relations between any two of the diffusion coefficient of 
				the
				quantities we construct, which are built through 
				statistical value
				of the lateral lengths.
				${D_V}\sim{\alpha_1}\langle{L_x^2}{L_z^4}\rangle{D_{B_1}}$,
				${D_V}\sim{\alpha_2}\langle{L_x^4}\rangle{D_{B_2}}$,
				${D_{B_1}}\sim{\alpha_3}\langle{L_x^4}/{L_y^4}\rangle{D_{B_2}}$,
				here $\alpha_1$, $\alpha_2$ and $\alpha_3$ are constants and
				numerically obtained from our simulations as: 0.0584,  
				0.3302 and 
				0.0195,	respectively  \label{fig4}}
		\end{figure}
		
		The numerical results are shown in the Fig. \ref{fig4}, the 
		diffusion
		coefficients (corresponding to the short time part in the
		Fig. \ref{fig2}(b)) are measured according to the Eq. \ref{eq3}, 
		the four
		data points in each plots are measured over $N=423, 1000, 1500$ and
		$2000$. Data is collected by averaging over 10 initial samples, each
		evolved 50 times over $10^7$ proposed bond transpositions. It is
		numerically found that
		${D_V}\sim{\alpha_1}\langle{L_x^2}{L_z^4}\rangle{D_{B_1}}$,
		${D_V}\sim{\alpha_2}\langle{L_x^6}\rangle{D_{B_2}}$,
		${D_{B_1}}\sim{\alpha_3}\langle{L_x^4}/{L_z^4}\rangle{D_{B_2}}$, the
		factors $\alpha_1$, $\alpha_2$ and $\alpha_3$ are constant and 
		fitted
		from our simulations: 0.0584, 0.3302 and 0.0195, respectively, 
		 we infer that these values are more likely to 
			depend on factors such as sample size, amount of data, 
			measurement time length, 
			system noise, etc., which can be used as references in future 
			experiments.

	\subsection{Characterizing material dynamics at long times\label{pro}}
	
	In comparison to the short time dynamics, the dynamics at long times
	is much more noisy, simply because obtaining very long time series for
	an a-Si sample is not easy. Just to give an idea, a run on a single
	core of Intel i7-9700k CPU, our simulations require 0.005s/step for a
	sample with 1000 atoms for $T=1500$K. Given that the crossover event
	often takes place at $10^4$ steps, at least $10^6$ steps, and
	averaging over 50 independent sample runs, costing five days, would be
	required for obtaining results with good statistics (larger samples
	and lower temperatures would require even longer times.) To complicate
	matters, we also have to deal with the stalling events. Fortunately,
	we have found a way to go around the complications due to the stalling
	events, as described in Sec. \ref{secpro1} below.

	\subsubsection{Stalling events complicate handling of time-series 
		data\label{secpro1}}
	
	\begin{figure}[H]
		\begin{center}
			\includegraphics[width=0.9\linewidth]{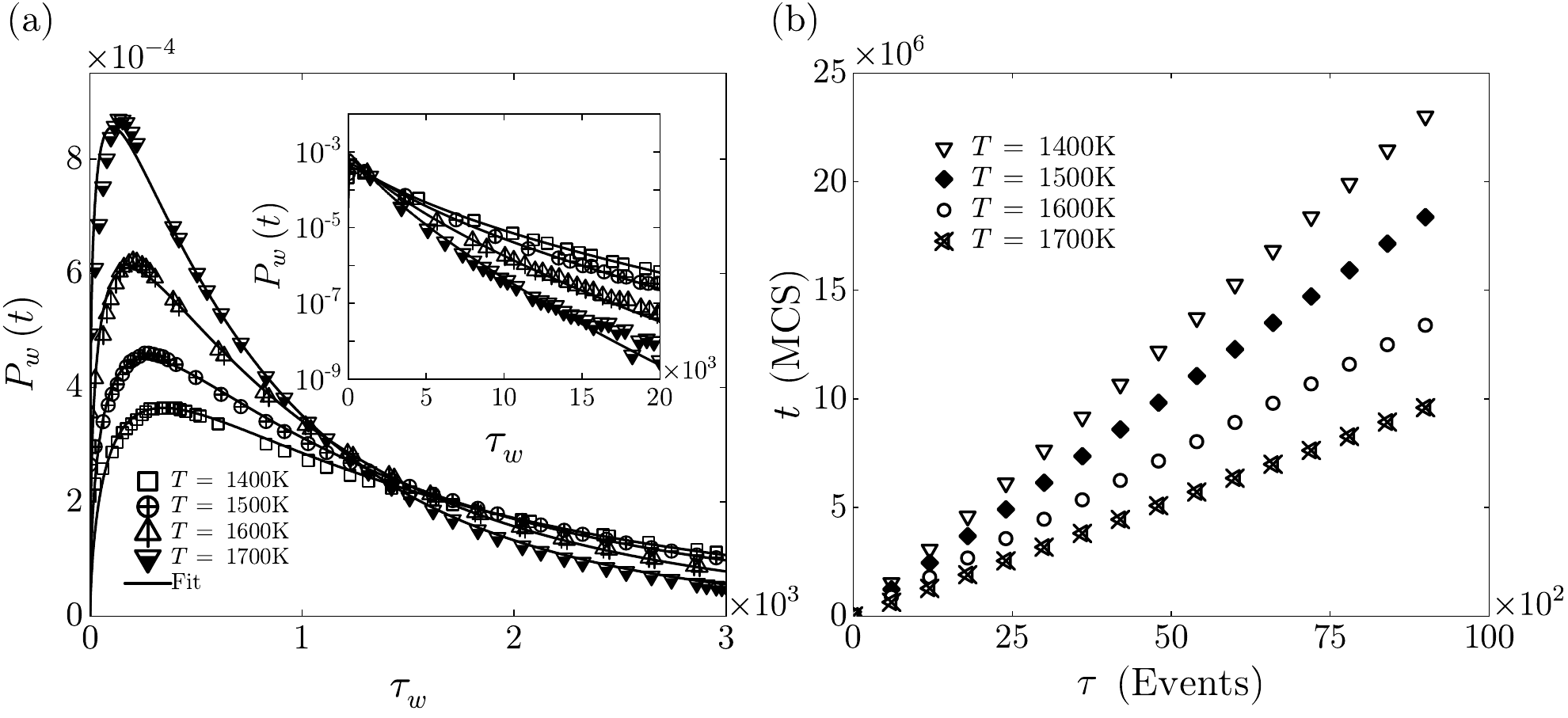}
		\end{center}
		\caption{  (a) Waiting time distribution fitted to
			Eq. (\ref{diswt}) for $N=423$ at various temperatures (up to 
			$\tau_w=$3,000), inset: fit for lager scales (up to 
			$\tau_w=$20,000). (b) Monte Carlo time $t$ for a sample with 
			$N=423$, measured in the number of proposed bond 
			transpositions, in 
			comparison with
			the event time $\tau$, measured in the number of accepted bond
			transpositions. Each dot in this scatter plot is a measurement 
			of
			both times, elapsed between accepted bond transpositions. The 
			data
			show a linear relation between $t$ and $\tau$, with a
			temperature-dependent scale factor. As is to be expected, there 
			is
			some spread in the data points at short times, which decreases 
			with
			increasing times.
			\label{fig5}}
	\end{figure}
	The stalling events pose us with a difficulty on how to cleanly handle
	the dynamics data at long times: specifically, $V$, $B_1$ and $B_2$
	changing only at irregular time intervals means that ensemble
	averages, calculated standardly from time-series data, is bound to be
	very noisy (it is!) at long times. Stalling events are a manifestation
	of the MC procedure: over a stalling period, which is synonymous to a
	{\it waiting time\/} (i.e., the sample configuration is waiting for a
	change), all proposed bond transposition events in the MC procedure
	are rejected. That said, the stalling events also provide us with the
	opportunity to measure time in units of {\it accepted\/} bond
	transposition moves, $\tau$, instead of the standard units of MC time
	$t$, measured in units of {\it attempted\/} bond transposition
	moves. Clearly, the quantity $\tau$ increases by unity at every
	accepted move, even though the MC time between two successive accepted
	moves, the waiting time can be widely different due to stochasticity
	of the process. Indeed, when measured in units of $\tau$, we find that
	the dynamical quantities at long times are far less noisy, and for
	this reason, throughout this section we measure time in units of
	$\tau$.
	
	But before we get to the dynamics, we present the probability
	distribution of the waiting time $P_w(t)$ in Fig. \ref{fig5}(a) for a
	sample with $N=423$ and a couple of different temperature
	values. Plotting the data in log-linear plot, and subsequent analysis,
	reveals that the $P_w(t)$ data are well fitted by the following formula:
	\begin{eqnarray}
		P_w(t) =at^be^{-(\tau_w/\tau_c) ^d},
		\label{diswt}
	\end{eqnarray}
	where $a$ is a normalization constant. The fitting parameters are
	noted in Table \ref{p_wt}. In particular, we note that the waiting
	time distributions do not have power-law tails. We will return to this
	aspect later in this section. Moreover, in Fig. \ref{fig5}(b) we also
	show that the mean waiting time is a constant throughout the duration
	of the simulation (the constant corresponds to the inverse rate of the
	accepted moves).

\begin{table}[H]
	\caption{\label{p_wt} Parameters for the fitted waiting time
		distribution in Fig. \ref{fig5}(a).}
	\centering
	\begin{tabular}{ccccc}
		\hline\hline
		\textbf{Temperature (K)} & $a$ & $b$ & $\tau_c$ & $d$ \\
		\hline
		1400 & 2.070 $\times 10^{-5}$ & 0.767 & 125.000 & 0.474 \\
		1500 & 4.721 $\times 10^{-5}$ & 0.615 & 195.100 & 0.525 \\
		1600 & 1.179 $\times 10^{-4}$ & 0.466 & 274.800 & 0.595 \\
		1700 & 3.315 $\times 10^{-4}$ & 0.300 & 350.000 & 0.667 \\
		\hline\hline
	\end{tabular}
\end{table}
	
	\subsubsection{Anomalous diffusion of $V$, $B_1$ and 
		$B_2$\label{secpro2}}
		\begin{figure}[H]
		\begin{center}
			\includegraphics[width=0.7\linewidth]{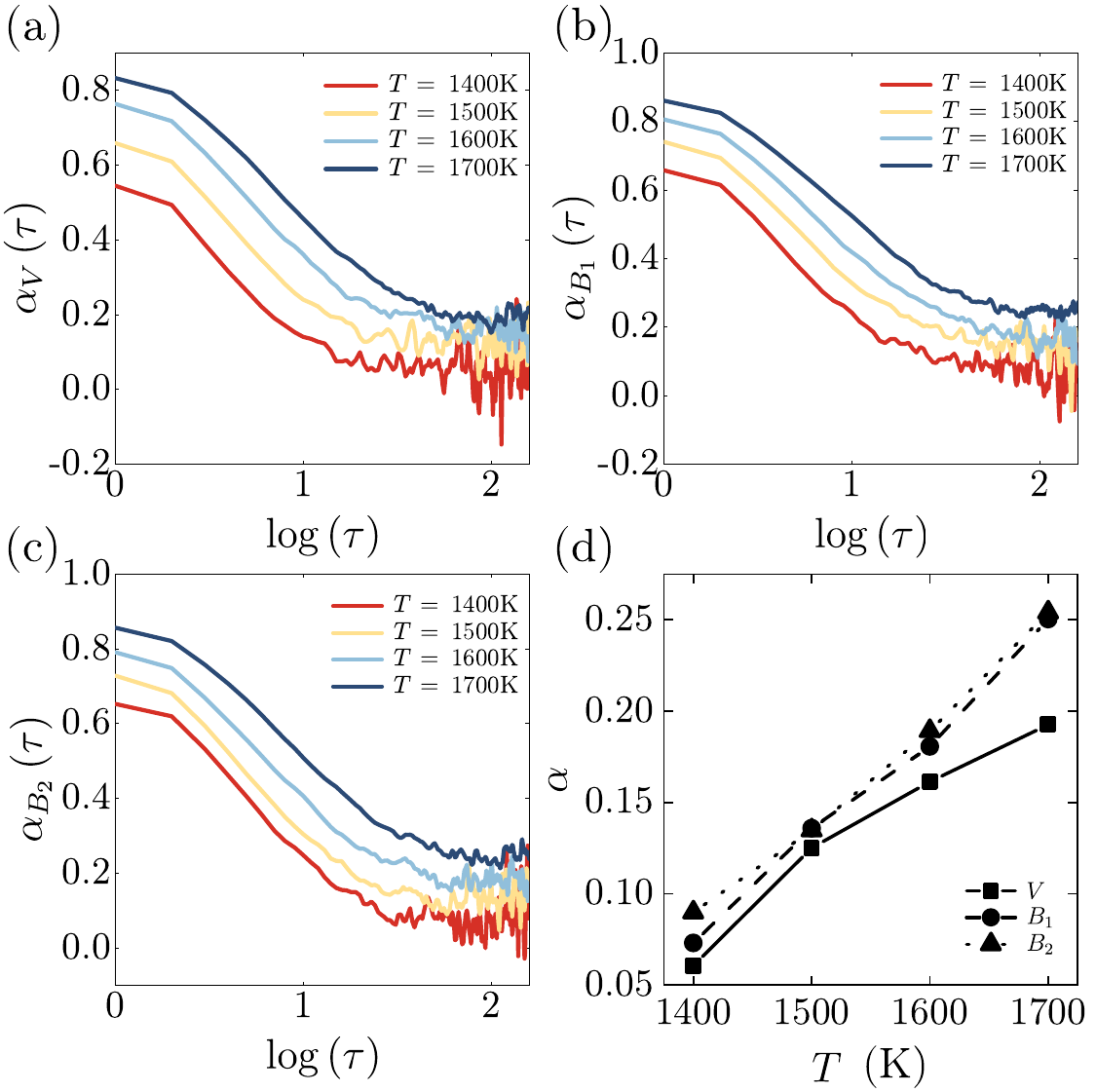}
		\end{center}
		\caption{(color online) (a)-(c) Measurement of the exponent 
			$\alpha$ in
			log($\tau$) scale for various temperatures with fixed $N=423$. 
			$\alpha$ is smaller then 1 throughout the entire time, which 
			indicates a
			subdiffusive behavior for the corresponding mean-square
			displacement. After an initial period of crossover time, 
			$\alpha$ eventually converges to a small value (0.1-0.3). 
			(d) The anomalous exponent (as defined in Eq. \ref{expo}) as obtained from the MSD of $B_1$, $B_2$ and $V$, as a 
			function of 
			temperature.  At higher temperatures, the values of $\alpha$ 
			obtained 
			from the MSD of $V$ are lower than those for $B_1$ and $B_2$, 
			probably 
			because large deviations of the density away from the 
			crystalline 
			value are suppressed (as discussed in the text).  
			\label{Talpha}}
	\end{figure}
	Double logarithmic plots of the mean-square displacements reveal that
	the dynamics of $V$, $B_1$ and $B_2$ is no longer diffusive, i.e.,
	they are anomalous, at long times. Assuming that the
	$\text{MSD}_V(\tau)$, $\text{MSD}_{B_1}(\tau)$ and
	$\text{MSD}_{B_2}(\tau)$ increase as a power-law in $\tau$ at long
	times, we compute the effective exponents for these variables,
	collectively denoted by $Q$, as
	\begin{equation}
		\alpha _Q\left( \tau \right) =\frac{d\log 
			\left[\text{MSD}_Q\right]}{d\log 
			\tau}.
		\label{expo}
	\end{equation}
	An example plot for $\alpha_Q$ is shown for $N=423$ in
	Fig. \ref{Talpha}(a)-(c) above. The exponents we find are dependent on
	temperature (shown in Fig. \ref{Talpha}(d)). It seems to us that lower
	temperature leads to lower subdiffusion exponent, even though
	measuring exact exponents properly is difficult given long run
	times. The approximate exponents for each quantities at various
	temperature are list in Tab. \ref{Tab_alpha}.

\begin{table}[H]
	\caption{\label{Tab_alpha} The anomalous diffusive exponent $\alpha$ for $V$, $B_1$, and $B_2$ with fixed $N=423$ and various $T$.}
	\centering
	\begin{tabular}{ccc}
		\hline\hline
		\textbf{Quantity} & \textbf{$T$ (K)} & $\boldsymbol{\alpha}$ \\
		\hline
		$V$ & 1400 & 0.06 \\
		$V$ & 1500 & 0.13 \\
		$V$ & 1600 & 0.16 \\
		$V$ & 1700 & 0.20 \\
		\hline
		$B_1$ & 1400 & 0.07 \\
		$B_1$ & 1500 & 0.14 \\
		$B_1$ & 1600 & 0.18 \\
		$B_1$ & 1700 & 0.25 \\
		\hline
		$B_2$ & 1400 & 0.09 \\
		$B_2$ & 1500 & 0.13 \\
		$B_2$ & 1600 & 0.19 \\
		$B_2$ & 1700 & 0.25 \\
		\hline\hline
	\end{tabular}
\end{table}

	\subsubsection{Further analysis of anomalous diffusion \label{secpro3}}
	
	Long runtimes required to obtain long time series of sample snapshots
	prevent us to pinpoint the exponents with higher numerical
	accuracy. Nevertheless, we can still reflect on the nature of the
	anomalous diffusion we observe here for a-Si. In particular, in the
	past work of two of us, we have observed that anomalous diffusion
	in materials tend to belong to the fractional Brownian motion (fBm)
	class, while their stochastic dynamics described by Generalized
	Langevin equation (GLE) with power-law memory 
	\cite{panja2010generalized,panja2010anomalous,zhong2018generalized,panja2015efficient},
	with the exponent
	of the power-law memory, within the numerical accuracy, matching the
	anomalous diffusion exponent really well. The memory can be
	interpreted in terms of restoring forces: when thermal fluctuations
	move a sample configuration one way, subsequent fluctuations tend to
	undo that move. If the simulations of a-Si were amenable to reach
	sufficiently long times, we would be able to perform a similar analysis
	here as well; unfortunately that is not the case.
	
	\begin{eqnarray}
		C_{v}^{Q}\left( \tau \right) =\frac{\left< v_Q\left( \tau \right) 
			v_Q\left( 0 
			\right) \right>}{\left< \left( v_Q\left( 0 \right) \right) ^2 
			\right>}
		\label{vacfeq}.
	\end{eqnarray}
	
	That said, given that the waiting times do not have a power-law tail,
	as demonstrated in Fig. \ref{fig5} rules out a continuous-time random 
	walk
	(CTRW) type stochastic process for $V$, $B_1$ and $B_2$. To completely
	rule out CTRW, we plot the velocity autocorrelation function (VACF)
	for these quantities in Fig. \ref{vacf}, VACF 
		$C_{v}^{Q}\left( \tau \right)$ is defined in Eq. \ref{vacfeq} for 
		quantity $Q$, 
		where $v_Q\left( \tau \right) =\partial Q/\partial \tau $ is the 
		velocity of 
		the quantity Q. These data rather cleanly demonstrate that the 
	VACF is 
	negative for $\tau\neq0$,
	and approaches zero for large $\tau$ from below the $x$-axis.
	
		\begin{figure}[H]
		\centering
		\includegraphics[width=0.5\linewidth]{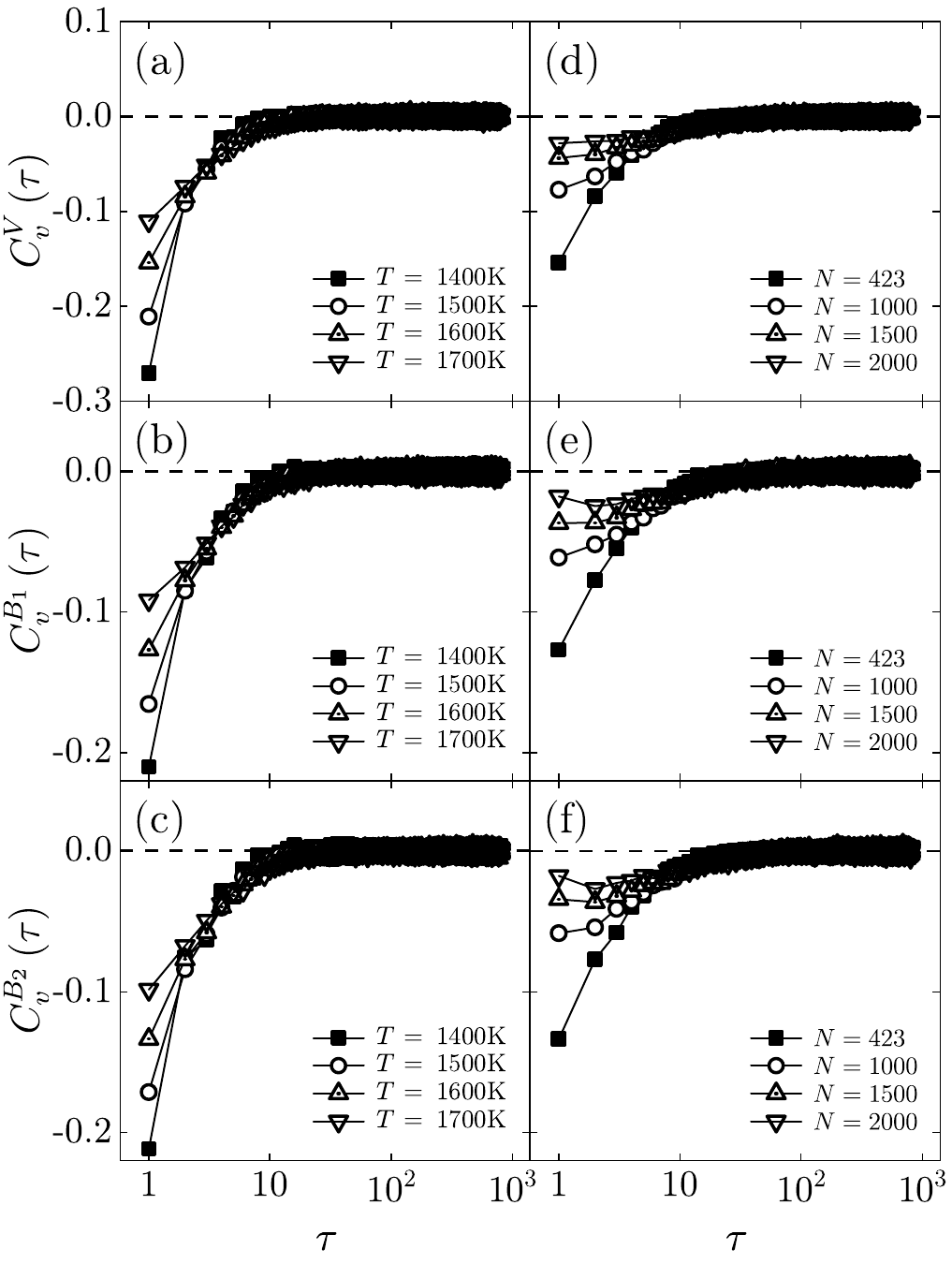}
		\caption{ Normalized velocity auto-correlation
			functions (VACF) of $V$, $B_1$ and $B_2$ for various 
			temperatures with 
			fixed $N=423$ (a)-(c) , and (d)-(f) for various the number of 
			atoms 
			with a fixed temperature of 1500K. The VACF are clearly 
			negative at 
			very short times,
			indicating that there is a restoring tendency: if in one bond
			transposition, the sample shears one way, the following bond
			transposition has a bias in favor of undoing the earlier shear.
			It also implies the autocorrelation is weaker for a larger
			system. \label{vacf}}
	\end{figure}  
	
	For completeness, we also present the jump length distributions of 
	$B_1$, $B_2$ and $V$
	in Fig. \ref{disjump}. Note that these distributions do not feature fat 
	tails, 
	thereby ruling out Levy-flight like effects.
	
	\section{Discussion \label{con}}    
    
    	\begin{figure}[H]
    	\includegraphics[width=0.8\linewidth]{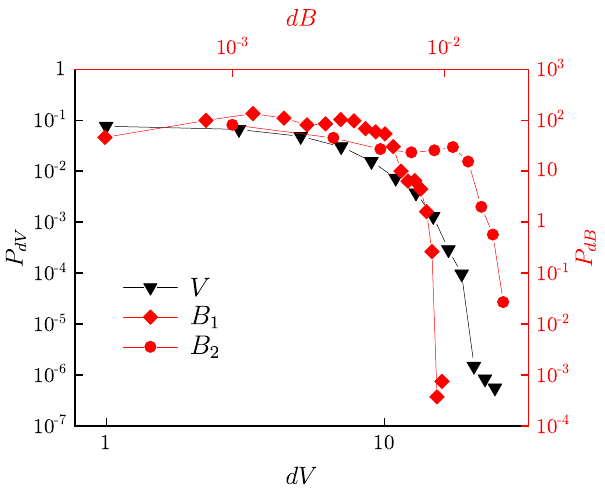}
    	\centering
    	\caption{(color online) Double logarithmic plots of the jump length 
    		distribution for $B_1$, $B_2$ 
    		and $V$ ($N=423$, $T=1500$K). None of these distributions 
    		feature fat 
    		(power-law) tails, thereby ruling out Levy-flight behavior.  
    		\label{disjump}}
    \end{figure}
    
	This manuscript studies the structural dynamics of a model of amorphous 
	silicon, in particular
	the fluctuations in the volume $V$ of the simulation box, as well as in 
	its aspect ratios $B_1$ and $B_2$.
	The simulations show that these three variables show ordinary diffusive 
	behavior at very short times,
	crossing over to anomalous diffusion at longer times: their MSDs can be 
	well fitted with a power-law
	MSD$\sim t^\alpha$ with $\alpha<1$. We find that the anomalous exponent 
	$\alpha$ is temperature-dependent.
	
	Various models in statistical physics exist that also feature anomalous 
	dynamics; two well-known examples are
	the continuous time random walker (CTRW) and fractional Brownian motion 
	(fBM). In further investigation,
	we present the distribution of the waiting times, as well as the 
	velocity autocorrelation functions. These
	further results show that the observed anomalous dynamics is consistent 
	with fBM-like behavior, and not with
	CTRW-like behavior.
	
	The observed fluctuations in the shape of the simulation cell are 
	directly related to deformations of the material
	under external compression and shear forces. Our findings are therefore 
	directly linked to experiment.

	
	\end{document}